\documentclass[journal,twocolumn, 10pt]{IEEEtran}

\usepackage{epsfig,makeidx,color,epstopdf}
\usepackage{amsmath,amssymb,amsthm,bbm}
\usepackage{cite,graphicx}
\usepackage{enumerate}
\usepackage{amsfonts}
\usepackage{mathtools, cuted}
\usepackage{lipsum}
\usepackage{indentfirst}
\usepackage{subfig}
\usepackage[font=footnotesize]{caption} 
\usepackage{csquotes}
\usepackage{float}
\usepackage{comment}

\usepackage{array}
\newcolumntype{P}[1]{>{\centering\arraybackslash}p{#1}}
\newcolumntype{M}[1]{>{\centering\arraybackslash}m{#1}}

\makeatletter
\def\footnoterule{\kern-3\p@
  \hrule \@width 2in \kern 2.6\p@} 
\makeatother

\def\defh{
\mbox{\footnotesize $ \begin{array}{c} H_0 \cr > \cr < \cr H_1 \end{array} $}}
\def\rdefh{
\mbox{\footnotesize $ \begin{array}{c} H_1 \cr > \cr < \cr H_0 \end{array} $}}

\def\be{ \begin{equation} }
\def\ee{ \end{equation} }
\def\bea{ \begin{eqnarray} }
\def\eea{ \end{eqnarray} }

\def\b0{{\bf 0}}

\def\cS{{\cal S}}

\ifCLASSOPTIONonecolumn
\else
  
\fi

\makeatletter
\newcommand*\dashline{\rotatebox[origin=c]{90}{$\dabar@\dabar@\dabar@$}}
\makeatother
\makeatletter
\newcommand*{\rom}[1]{\expandafter\@slowromancap\romannumeral #1@}
\makeatother

\begin{document}
\title{Short-Range Ambient Backscatter Communication Using Reconfigurable Intelligent Surfaces} 
\author{\IEEEauthorblockN{Mahyar Nemati, Jie Ding and Jinho Choi}\\
        \IEEEauthorblockA{School of Information Technology, Deakin University, Melbourne, VIC 3125, Australia \\
        e-mail: nematim@deakin.edu.au, j.ding@deakin.edu.au,  jinho.choi@deakin.edu.au
        }
        }
        
%
\date{today}
\maketitle
\pagenumbering{gobble} 

\begin{abstract}
Ambient backscatter communication (AmBC) has been introduced to address communication and power efficiency issues for short-range and low-power Internet-of-Things (IoT) applications. On the other hand, reconfigurable intelligent surface (RIS) has been recently proposed as a promising approach that can control the propagation environment especially in indoor communication environments.
In this paper, we propose a new AmBC model over ambient orthogonal-frequency-division-multiplexing (OFDM) subcarriers in the frequency domain in conjunction with RIS for short-range communication scenarios. A tag transmits one bit per each OFDM 
subcarrier broadcasted from a WiFi access point. Then, RIS augments the signal quality at a reader by compensating the phase distortion effect of multipath channel on the incident signal. We also exploit the special spectrum structure of OFDM to transmit more data over its squeezed orthogonal subcarriers in the frequency domain. Consequently, the proposed method improves the bit-error-rate (BER) performance and provides a higher data rate compared to existing AmBC methods. Analytical and numerical evaluations show the superior performance of the proposed approach in terms of BER and data rate. 
\end{abstract}
{\IEEEkeywords
AmBC, OFDM, intelligent surface, reflector.}

\ifCLASSOPTIONonecolumn
\baselineskip 26pt
\fi

\section{Introduction}
Ambient backscatter communication (AmBC) has been actively studied for future short-range and low-power Internet-of-Things (IoT) applications \cite{van2018}. This new battery-less technology enables a low-power tag to sense and modulate ambient radio-frequency (RF) signals and send its own data to nearby devices with the same spectrum utilization. In general, an AmBC system includes a tag, a reader, an ambient RF source, and a legacy receiver. The source broadcasts its RF signal to communicate with the legacy receiver. The tag also receives the RF signal through the forward-link channel due to its relatively short distance from the source. Different from the conventional backscatter communications, such as in radio identification (RFID) systems \cite{RFIDboyer,rfid,rev}, AmBC does not require a dedicated source for direct device-to-device (D2D) and even multi-hop communications \cite{RFIDint,of-3}. An integrated circuit of the tag contains components to harvest/capture enough power from the ambient RF signal; and then, modulate and backscatter it towards the reader. The reader also receives the original ambient RF signal through the direct-link channel. Since, the RF signal is unknown to the reader, it is considered to be direct-link interference for transmitted data from the tag. Nevertheless, due to this inteference and power constraints, AmBC is suitable for short-range communications only.

Besides, reconfigurable intelligent surfaces/walls/arrays (RIS) have been recently proposed as a promising technology to control the propagation environment \cite{IS1,IS3,IS2}. RIS includes a large number of simple low-cost reflectors to modify the phase of incident signals and reflect them to an intended receiver. This appealing technology can cause a major performance improvement of communications in rich-scattering environments especially in indoor communication scenarios. 
Contrary to the AmBC, RIS does not change the modulation of ambient signals. It only adjusts a phase-shift in the incident signal to compensate the phase distortion effect of the multipath channels \cite{IS2}. Consequently, RIS controls the signal propagation to enhance the quality of the signal at the receiver.  
As a brief description for both of AmBC and RIS technologies and in order to clarify them, we have Table \ref{sum} that provides some important characteristics of both.

\begin{table}[t]
\centering
	\caption{Brief comparison between AmBC and RIS technologies.}\label{sum}
	\begin{tabular}{|m{1.5cm}|m{2.5cm} | m{3cm}|}
	\hline
	& \hspace{.8cm}AmBC & \hspace{1.2cm} RIS\\
	\hline
	\hline
	Purpose & Information transmission over ambient RF signals. & Signal quality enhancement by controlling the propagation environment.\\
	\hline
	Main component & A backscatter device. &  A large number of reflectors on a surface.  \\
	\hline
	Effect on incident signal & Modulating the incident signal and waveform modification. &  Only reflects the incident signal with an adjustable phase-shift.\footnotemark \\
	\hline
	Energy source & No dedicated energy source. & No dedicated energy source.\\
	\hline
	Coding & Encoding (and rarely decoding). &  No decoding/no encoding.\\
	\hline
	\end{tabular}
\end{table}
\footnotetext{In some cases, RIS can also amplify the amplitude of the incident signal \cite{Amp1}.
Therefore, a) continuous amplitude and phase-shift, b) constant amplitude and continuous phase-shift, and c) constant amplitude and discrete phase-shift are three different assumptions for RIS reflectors coefficients \cite{LISA}. }

Since, RIS can enhance the performance of communications in rich-scattering environments and also AmBC is an appropriate connectivity technology for low-power applications, we aim to combine these two technologies for an efficient short-range communication in indoor environments which can be used for \textit{smart home applications}. Therefore, we focus on WiFi signals, with major focus on the special structure of orthogonal-frequency-division-multiplexing (OFDM), which are ubiquitous signals in current and future smart homes.

In the literature, there is a significant effort to apply AmBC over OFDM signals. AmBC was firstly set up over ambient digital TV (DTV) signals for a very short-range (up to $1$m) D2D communication with a data rate of $1$kbps \cite{liu2013}. Likewise, a relatively similar AmBC system in \cite{chan2} achieved a higher data rate of $20$kbps using WiFi signals for a range of $2$m. Subsequently, a data rate of $1$Mbps was obtained in \cite{backfi} by using the principles of full-duplex systems \cite{of-3}. Afterwards, a noncoherent detection method using multiple antennas was proposed in \cite{wang2016} to remove the need of channel state information (CSI) for signal detection in AmBC systems. 
 However, none of the mentioned articles evaluated or exploited the physical layer structure of the OFDM signals broadcasted from WiFi access points for performance improvement. 
OFDM waveform has a special spectrum structure which can provide a range of flexibility in terms of data rate. In \cite{of-0,of,Sensor,Jin_OF}, two different AmBC methods were proposed to transmit only one bit over one OFDM symbol (duration). 
 In \cite{of-0,of,Sensor}, an AmBC approach was investigated which exploits the cyclic prefix (CP) of OFDM signals to improve the performance of the system when the duration of the CP is greater than the maximum delay spread of multipath channels. A method in \cite{Jin_OF} used matched-filtering at the tag to impose a certain property that improves the signal detection performance at the reader. Furthermore, in \cite{of1,F-Guanding}, an AmBC method was evaluated to send data over null (i.e., guard-band) subcarriers of an OFDM signal. However, they either increase the complexity of the tag like in \cite{Jin_OF,of1,F-Guanding}, or their bit-error-rate (BER) performance and data rate are not satisfactory yet like in \cite{of-0,of,Sensor}. 

In this paper, we propose a new AmBC technique over ambient OFDM subcarriers in the frequency domain in conjunction with the RIS for the first time which is the main contribution of our study. We exploit the special spectrum structure of OFDM to transmit a higher data rate over its squeezed orthogonal subcarriers in the frequency domain. To be more specific, a tag transmits one bit per each ambient OFDM subcarrier broadcasted from a WiFi access point; and RIS augments  the quality of the signal at a reader by compensating the phase distortion effect of multipath channel on the incident signal. At the tag, on-off keying (OOK) modulation is employed to nullify an ambient subcarrier to transmit bit $0$ or maintain the ambient subcarrier to transmit bit $1$. 
The major aims of the proposed model are to achieve a higher transmission rate and a superior BER performance for short-range backscatter communication over ambient OFDM. Furthermore, the method can take advantage of index modulation (IM) technique \cite{IM} to control the signal power and reduce the interference imposed on original OFDM signal.

 \textit{Notation:} $(.)^T$ and $\lfloor .\rfloor$ denote the transpose operation and the floor operation, respectively. The 2-norm of $\mathbf{a}$ is denoted by $||\mathbf{a}||$. $\mathcal{CN} (\mathbf{a},\mathbf{R})$ represents the distribution of circularly symmetric complex Gaussian (CSCG) rendom vectors with mean vector $\mathbf{a}$ and covariance matrix $\mathbf{R}$. The Gaussian Q-function is given by $\mathcal{Q}(x)=\frac{1}{\sqrt{2\pi}}\int^\infty_x e^{\frac{-z^2}{2}} dz$. 
\section{System Model}
\begin{figure}[t]
\centering
\captionsetup{width=1\linewidth}
  \includegraphics[width=1\linewidth, height=5.1cm]{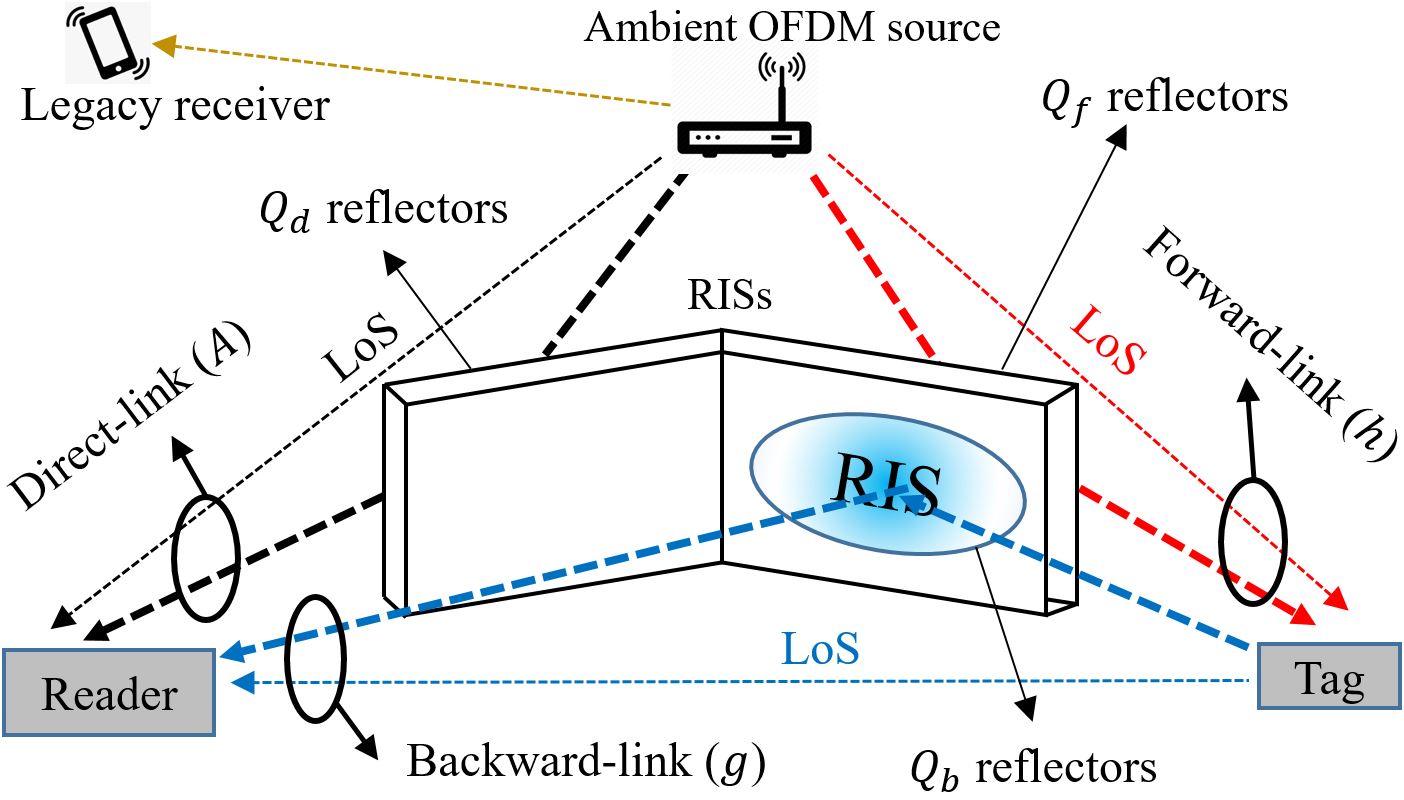}%
  \caption{Illustration of AmBC system model in coexistence with RIS. There are three groups of $Q_f$, $Q_d$, and $Q_b$ reflectors on different sides of the wall for forward-link, direct-link, and backward-link channels, respectively. }%
  \label{fig:sys}
\end{figure}
The system model of the new proposed approach is shown in Fig. \ref{fig:sys} and described in three stages as follows.

At the first stage, at the source, the OFDM symbol with a total of $N_{s}$ subcarriers is transmitted, which is given by
\begin{equation}
\mathbf{x}=
\mathbf{U}\, \mathbf{s},
 \label{eq11}
\end{equation}
where $\mathbf{s}=[s_1,\, s_2,\, \dots, s_{N_s}]^T$ 
 is the vector of modulated data symbols. We omit the guard-band subcarriers for notational simplicity to avoid any confusion since it does not affect our system. 
 Here, $\mathbf{U}\triangleq[ \mathbf{M}^T \, \mathbf{F}_{N_s}^{-1}]^T$ where $\mathbf{F}_{N_s}$ is the $N_s\times N_s$ fast Fourier transform (FFT) matrix and $\mathbf{M}$ represents the last $N_{cp}$ rows of the $\mathbf{F}_{N_s}^{-1}$ for CP \cite{jinB1}.%
 
We assume the ambient OFDM source, i.e., WiFi access point in an indoor scenario, the tag, and the reader are in a short distance from each other like in smart home applications. As shown in Fig. \ref{fig:sys}, there is a RIS wall with a large number of simple low-cost reflectors in the middle of the wireless channel. The signal received at the tag is the
superposition of the line-of-sight (LoS) signal and the signals reflected by
RIS reflectors. It strongly depends on
the phases of the multipath propagation \cite{IS3} and can be represented by

\begin{equation}
    \mathbf{u}=\left( a_0 e^{-j\theta_0} + \sum_{i=1}^{Q_f} a_i e^{-j(\theta_i-\psi_i}) \right)    \, \mathbf{x},
    \label{eq_IS1}
\end{equation}
where $a_i$, $\theta_i$, and $\psi_i$ are the attenuation of the $i^{\text{th}}$ path, phase-shift of the $i^{\text{th}}$ path, and the phase induced by the $i^{\text{th}}$
reflector, respectively. Note that the $0^{\text{th}}$ path corresponds to
the LoS path. $Q_f\in \mathbb{N}$ denotes the total number of reflectors of the RIS for the forward-link channel as shown in Fig. \ref{fig:sys}. In order to maximize the signal power and consequently increase the signal quality at the tag, vector $\boldsymbol{\psi}=[ \psi_1,\, \psi_2,\cdots \psi_{Q_f}]^T$ should be adjusted in a way to compensate the phase distortion effect of the multipath channel, i.e., $[\boldsymbol{\theta}]_i-[\boldsymbol{\psi}]_i=\theta_0$, $i=1,...,Q_f$. Consequently, (\ref{eq_IS1}) can be written as
\begin{equation}
    \mathbf{u}=\left( a_0 + \sum_{i=1}^{Q_f} a_i\right) e^{-j\theta_0}   \, \mathbf{x} = h\, \mathbf{x}.
    \label{eqIS2}
\end{equation}

 In fact, as the number of reflectors on RIS increases, the quality of the signal at the tag increases. In other words, RIS modifies the multipath forward-link channel and enhances the system performance in rich-scattering environments.
Moreover, the maximum delay spread is less than the sampling period due to the short distance scenario. Thus, the paths are considered to be one resolvable path with strong LoS. Therefore, the forward channel from the source to the tag, i.e., $h$, is assumed to have a single path time-variant channel coefficient. In addition, the channel coherence time is considered to be greater than one OFDM symbol period due to low mobility of devices with respect to each other. As a result, flat fading channels can be assumed during one OFDM symbol period. Likewise, the direct-link channel from the source to the reader profits from $Q_d\in \mathbb{N}$ reflectors on the other side of the RIS, as shown in Fig. \ref{fig:sys}, and acts as a flat fading channel with $A$ as its single path time-variant channel coefficient, i.e., $A=\sum_{i=0}^{Q_d} b_i e^{-j\varphi_0}$ where $b_i$ is the attenuation of the $i^{\text{th}}$ path; and $\varphi_0$ refers to the phase-shift of the LoS path.
Thus, the received OFDM symbol at the reader
becomes
\begin{align}
& \mathbf{y}_d=A \, \mathbf{x}.
\label{eq5}
\end{align}


At the second stage, at the tag, the desired data is a vector $\mathbf{d}=[d_1, \, d_2,\, \dots, \, d_{N_s}]^T$, where $d_l\,( l=1,...,\, N_s) $ represents the data symbol of the tag transmitted through the $l^{th}$ subcarrier.
If OOK is employed, we can have $d_l \in \{0, 1\}$.
Throughout the paper, we assume that $d_l$ is regarded as a bit
for convenience.
 Letting $\mathcal{O}_{N_s\times N_{cp}}$ be an $N_s\times N_{cp}$ zero matrix, define 
\begin{equation}
\mathbf{V}_b\triangleq
\mathbf{U}\,
  ( \text{diag}(\mathbf{d}))\, \left[ \mathcal{O}_{N_s\times N_{cp}} \, \mathbf{F}_{N_s}\right].
\label{eq3}
\end{equation}

Then, the tag backscatters the multiplication of $\mathbf{V}_b$ by $\mathbf{u}$ to the reader. 
It nullifies the subcarriers corresponding to $d_l=0$ and preserves the subcarriers corresponding to $d_l=1$ in the frequency domain. Additionally, the backscattered noise is particularly negligible as compared to the reader noise which is a common supported assumption in \cite{van2018,LISA,IS4} and references therein. 
However, the signal is attenuated by a factor of $\beta$ inside the tag. 
The proposed method slightly increases the the power-consumption of the tag due to FFT/inverse FFT (IFFT) operations. However, an implementation of small sizes new FFT and IFFT blocks at the tag is not as challenging as before thanks to the new power saving mechanisms which can support the power of modern passive/semi-passive RFID tags and increase their battery life-time significantly, e.g., power saving mode (PSM) and expanded discontinuous reception (eDRX) mechanisms in \cite{7}. Nevertheless, our method is not restricted to only passive tags.

The backward channel also depends on the phases of the backward-link multipath propagation. Therefore, similar to the effect of the RIS for the forward-link channel in (\ref{eqIS2}), the backward-link channel utilizes $Q_b\in \mathbb{N}$ reflectors on the third side of the RIS and is being considered a single path time-variant coefficient which is denoted by $g$ as shown in Fig. \ref{fig:sys}, i.e., $g=\sum_{i=0}^{Q_b} c_i\, e^{-j\phi_0}$ where $c_i$ is the attenuation of the $i^{\text{th}}$ path and $\phi_0$ is the phase-shift of LoS channel. Thus, the received backscattered signal at the reader is given by
\begin{align}
\mathbf{y}_b
&= g\beta \mathbf{V}_b \mathbf{u}=\underbrace{g\,\beta\, h\,}_\alpha \mathbf{V}_b \, \mathbf{x},
\label{eq4}
\end{align}
where $\alpha$ denotes the overall composite source-tag-reader channel gain which is usually referred as the dyadic-backscatter-link in the literature \cite{LISA}.


Finally, at the third stage, from (\ref{eq5}) and (\ref{eq4}), the received signal with the background noise $\mathbf{n}\sim \mathcal{CN}(0,N_0)$ at the reader is given by
\begin{equation}
\mathbf{z}=\mathbf{y}_d+\mathbf{y}_b+\mathbf{n}=(A+\alpha \mathbf{V}_b)\mathbf{x}+\mathbf{n}.
\label{eqs}
\end{equation}

It is worth noting that since the length of the OFDM symbol is much longer than the propagation delay due to short distance between the tag and the reader, we ignore the propagation delay \cite{van2018}. %
%
At the reader side, the CP portion is discarded and the remaining part, denoted by vector $\mathbf{z}_d$, goes through the FFT operation (i.e.,  $\mathbf{F}_{N_s} \, \mathbf{z}_d$). Let $\mathbf{r}=[r_1,\, r_2,\,\dots r_{N_s} ]^T$ represent the received signal over the $N_s$ used subcarriers in the frequency domain, where $r_l$ is the received signal through the $l^{th}$ subcarrier.
%
From (\ref{eqs}), $r_l$ is written as
\be
r_l = (A + \alpha d_l) s_l + n_l,\quad \ l = 1,\ldots, N_s.
	\label{EQ:r2}
\ee

In the following subsections, we discuss three different features of the proposed method.
\vspace{-3mm}
\subsection{Discussion on Data Transmission Rate} Contrary to the existing methods in \cite{of-0,of,Sensor} which transmit one bit over one OFDM symbol duration, our proposed method can transmit up to $N_s$ bits over one OFDM symbol duration. To be more specific, considering an OFDM system based on 802.11a specifications (i.e., OFDM symbol duration of $T=4\mu$sec, FFT size of $64$, and $N_s=52$ used subcarriers), the existing methods can support a data rate of up to $\eta=\frac{1}{4\mu\text{sec}}=250$kbps; while our proposed method can support a data rate of $\eta=\frac{52}{4\mu\text{sec}}= 13$Mbps. 
Such a performance is up to $N_s$ orders of magnitude better than the existing AmBC systems over OFDM signals in the literature. 

\vspace{-1mm}
\subsection{Discussion on Coverage Enhancement} It is noteworthy that in the existing methods \cite{liu2013,chan2,backfi,Jin_OF,of-0,of,of1,F-Guanding,Sensor,wang2016}, the tag and the reader should be located in a very short distance from each other to avoid severe phase distortion of the channel due to the multipath propagation. However, our method can increase the range of the communication by utilizing RIS. 
For example, those existing methods profit from a strong LoS backward-link channel with a channel gain of $c_0$; while, our proposed method utilizes the set of multipath channel gains of $\sum_{i=0}^{Q_b} c_i$. Since $c_i \in \mathbb{R}^{+}$, it is evident that $\sum_{i=0}^{Q_b} c_i \gg c_0$. To be more specific, assuming that the pathloss increases with the square of the distance, denoted by $\mathsf{d}$, the ratio of the coverage in our approach compared to the existing methods can be modeled as
\begin{equation}
    \left(\frac{\mathsf{d}_{\text{new}}}{\mathsf{d}_{\text{old}}}\right)^2=\frac{\mathbb{E}\left\{\left(\sum_{i=0}^{Q_b} c_i\right)^2\right\}}{\mathbb{E}\{c_0^2\}}=\frac{Q_b^2\mu_c^2+Q_b\sigma_c^2}{c_0^2},
    \label{eqdc}
\end{equation}
where $\mu_c$ and $\sigma_c^2$ denote the mean and variance of $c_i,\, i=0,...,Q_b$. This coverage enhancement can also be extended to the forward-link and direct-lick channels. However, it is worth to mentioning that the new coverage is still restricted by the propagation delay between the backscattered signal, i.e., $\mathbf{y}_b$, and direct-link signal, i.e., $\mathbf{y}_d$, that has to remain negligible compared to the OFDM symbol duration which can be guaranteed in short-range indoor environments.
\vspace{-3mm}
\subsection{Discussion on Signal Power and Interference Level}
Utilizing OOK modulation, $N_s$ bits can be transmitted by nullifying or preserving ambient OFDM subcarriers. With equally likely $d_l$ (i.e. $\Pr(d_l = 0) = \Pr(d_l = 1) = \frac{1}{2}$) in OOK, the power of the backscattered signal changes randomly. Besides, there are an average of $N_s/2$ active subcarriers being backscattered which become interfering signals to a legacy receiver. Therefore, to control the power of the backscattered signal and keep the interference level low, it would be desirable to use a fixed fraction of $N_s$ subcarriers, say $K(\ll N_s/2)$ active subcarriers and $M(=N_s-K)$ null subcarriers for backscatter communication. To this end, we can consider IM technique \cite{IM}. The number of bits to be transmitted by IM is
\begin{equation}
\eta=\lfloor \log_2 \binom{N_s}{K} 
\rfloor.
\label{eqeta}
\end{equation}
%

 Consequently, the power of the signal, denoted by $E_s$, in relation to the power of the data symbol, $s_l$, denoted by $E_b$, is given by
\begin{align}
E_s=&\left(M(A)^2+ \,K(A+\alpha)^2\right)\frac{E_b}{N_s}.
\end{align}
\vspace{-4mm}
\section{Signal Detection and BER Analysis}
\subsection{Signal Detection} 
At the reader, 
the conditional distribution of $r_l$ on $s_l$ and $d_l$
is given by
\be
f(r_l\,|\, d_l, s_l) 
= \frac{1}{\sqrt{\pi N_0}} \exp
\left(
-\frac{1}{N_0}| r_l - (A + \alpha d_l) s_l|^2
\right).
\ee
Provided that $d_l$ is equally likely
(i.e., $\Pr(d_l = 0) = \Pr(d_l = 1) = \frac{1}{2}$),
the maximum likelihood (ML) detection is equivalent to the maximum posteriori probability (MAP) detection \cite{jinB2}, which is given by
\be
f(r_l \,|\, d_l = 0, s_l) 
\defh f(r_l \,|\, d_l = 1, s_l) ,
\ee
or
$\alpha s_l q_l \rdefh \frac{|\alpha  s_l|^2}{2}$
, where
$q_l = r_l - A s_l = \alpha s_l d_l + n_l$.

Since $s_l$ is unknown and regarded as a random variable,
the likelihood function of $d_l$ for given $r_l$
becomes
\be
f(r_l\,|\, d_l) = \sum_{s_l \in \cS} f(r_l\,|\, s_l, d_l) \Pr(s_l).
\ee
The ML detection becomes
\be
\sum_{s_l \in  \cS} 
e^{- \frac{1}{N_0} |r_l - A s_l|^2} \defh
\sum_{s_l \in  \cS} 
e^{- \frac{1}{N_0} |r_l - (A+\alpha) s_l|^2}.
	\label{EQ:Etest}
\ee
Note that this method is also applicable to modulated symbols with only one level of power. Suppose that
$
\cS = \{\pm \sqrt{E_b} \},$ and $s_l$ is equally likely (i.e., binary phase shift keying (BPSK)).
After some manipulations, 
the ML detection in \eqref{EQ:Etest} becomes
\be
\frac{\cosh \left( \frac{2 A \sqrt{E_b} r_l} {N_0} \right) }
{\cosh \left( \frac{2 (A + \alpha) \sqrt{E_b} r_l} {N_0} \right) }
\defh \exp
\left( - \frac{(2 A \alpha + \alpha^2) E_b}{N_0}
\right).
\ee
At a  high signal to noise ratio (SNR), since
\be
\frac{\cosh \left( \frac{2 A \sqrt{E_b} r_l} {N_0} \right) }
{\cosh \left( \frac{2 (A + \alpha) \sqrt{E_b} r_l} {N_0} \right) }
\approx
\exp \left( - \frac{2 \alpha \sqrt{E_b} |r_l|}{N_0} \right),
\ee
we have
\be
|r_l| \rdefh \frac{(2 A + \alpha) \sqrt{E_b}}{2}.
\label{ep}
\ee

Note that $\delta=\frac{(2A+\alpha)\sqrt{E_b}}{2}$ is the optimal decision threshold, i.e., $d_l=1$ if $|r_l|>\delta$ and $d_l=0$ if $|r_l|<\delta$.\\
\begin{figure}[t]
\centering
\captionsetup{width=1\linewidth}
  \includegraphics[width=5.8cm, height=3cm]{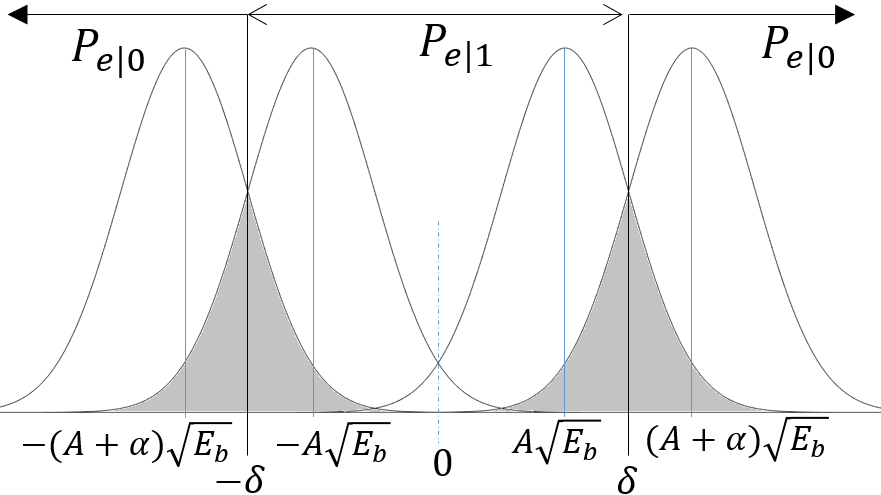}%
  \caption{The likelihood functions of the proposed method and BER regions for $P_{e|0}$ and $P_{e|1}$.}%
  \label{fig:pdf}
\end{figure}
\vspace{-3mm}
\subsection{BER Analysis} The reader can see four different subcarrier amplitudes from the received signal. Fig. \ref{fig:pdf} shows the likelihood functions of possible amplitudes of each subcarrier in the frequency domain, where the shaded areas represent error probabilities. Therefore, the error probability when bit 0 is transmitted is given by

\begin{align}
P_{e|0} & =\frac{1}{2\sqrt{\pi N_0}}\biggl[ \int_{-\infty}^{-\delta} e^{-\frac{(x+A\sqrt{E_b})^2}{N_0}} dx + 
                                                                      \int^{\infty}_\delta e^{-\frac{(x+A\sqrt{E_b})^2}{N_0}} dx  \nonumber\\
              & \,  \, +                                                  \int_{-\infty}^{-\delta} e^{-\frac{(x-A\sqrt{E_b})^2}{N_0}} dx+
                                                                      \int^{\infty}_\delta e^{-\frac{(x-A\sqrt{E_b})^2}{N_0}} dx \biggr].
\label{pe0}
\end{align}
Substituting $\delta$ from (\ref{ep}) into (\ref{pe0}), we have
\begin{align}
P_{e|0} = \mathcal{Q}\biggl(\sqrt{\frac{\alpha^2}{2}\gamma}\biggr) + \mathcal{Q}\biggl(\sqrt{2(2A+\frac{\alpha}{2})^2 \gamma}\biggr),
\end{align}
where $\gamma=\frac{E_b}{N_0}$ denotes the SNR. In a similar way,
\begin{align}
P_{e|1}=1-\mathcal{Q}\biggl(-\sqrt{\frac{\alpha^2}{2}\gamma}\biggr)-\mathcal{Q}\biggl(\sqrt{2(2A+\frac{3}{2}\alpha)^2\gamma}\biggr).
\end{align}
Finally, the BER is expressed as
\begin{align}
P_e&=\mathcal{Q}\biggl(\sqrt{\frac{\alpha^2}{2}\gamma}\biggr)+\frac{1}{2}\mathcal{Q}\biggl(\sqrt{2(2A+\frac{\alpha}{2})^2\gamma}\biggr)\nonumber\\
&- \frac{1}{2}\mathcal{Q}\biggl(\sqrt{2(2A+\frac{3}{2}\alpha)^2\gamma}\biggr)  \approx \mathcal{Q}\biggl( \sqrt{\frac{\alpha^2}{2}\gamma}\biggr).
\end{align}


\section{Numerical Results and Evaluations}
In this section, first, the BER performance of the system is evaluated. Second, the coverage performance evaluation is provided. Then, the data transmission rate along with the interference due to backscatter communication is studied. 

For the simulations, the parameters are considered as follows: $N_s=64$, $N_{cp}=16$, $T_d=3.2\, \mu s$, $T_{cp}=0.8\, \mu s$, $\beta=1$, $\sqrt{E_b}=1$ (e.g., BPSK), unless otherwise specified.\\
\vspace{-6mm}
\subsection{BER Performance Evaluation} 
\begin{figure}[t]
\centering
\captionsetup{width=1\linewidth}
  \includegraphics[width=9cm, height=6cm]{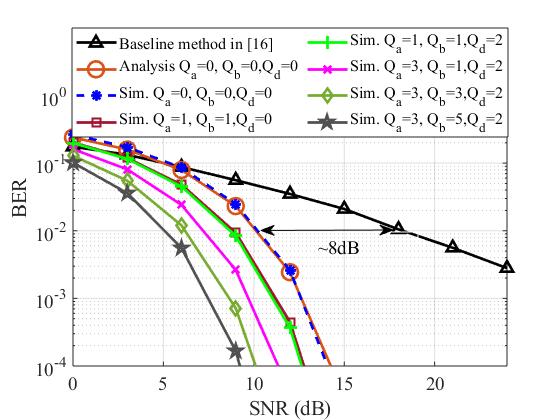}%
  \caption{BER evaluation of the proposed method.}%
  \label{fig:2}
\end{figure}
We compare the BER performance of our proposed method with the existing method in \cite{of-0} which is the baseline method for other existing methods in \cite{of,of1,F-Guanding,Sensor}. The BER performance of our proposed method depends on $\rho=\sqrt{\frac{\alpha^2}{2}\gamma}$. 
On the other hand, in flat fading channel and high SNR, the BER of the method in [16, eq. (27)] depends on a nonlinear function of $\gamma$, which can be expressed by
\begin{equation}
\zeta\approx \sqrt{\frac{N_{cp}(\gamma^2+(1-\sqrt{1+\frac{2\ln (\gamma)}{N_{cp}}})\gamma)^2}{\gamma^4}+2\ln (\gamma)}.
\label{eqsi}
\end{equation}

 As SNR increases in both the methods, $\rho$ gets bigger than $\zeta$ (i.e., $\rho>\zeta$). In addition, from (\ref{eqsi}), we can see that as $\gamma \rightarrow \infty$ then $\zeta^2 \rightarrow N_{cp}+2 \ln (\gamma)$, which means that the BER decreases slowly with $\gamma$. Since, $\mathcal{Q}(x)\leqslant \exp(-\frac{x^2}{2})$, as $\gamma \rightarrow \infty$, the BER of the method in \cite{of-0} becomes BER$\approx C \, \exp(-\ln (\gamma))=\frac{C}{\gamma}$, where $C$ is a constant. On the other hand, the BER of the proposed approach is BER$\approx \exp(-\frac{\alpha}{2}\gamma)$. Consequently, the BER in our proposed method outperforms the method in \cite{of-0}. 
 Fig. \ref{fig:2} shows that 
 our proposed method outperfoms the baseline method in \cite{of-0} in flat fading channels for a different number of reflectors on the RISs, e.g., around 8 dB SNR gain at BER$=10^{-2}$ without utilizing RIS reflectors when there is only strong LoS paths and $|A|=|\alpha|=1$ . 
 
 Furthermore, from Fig. \ref{fig:2}, we also conclude that the effect of $Q_d$ is almost negligible while $Q_f$ and $Q_b$ affect the BER performance significantly. In other words, when more reflectors are being deployed in RIS at the middle of dyadic-backscatter-link channel, i.e., larger $Q_f$ and $Q_b$, $|\alpha|$ in (\ref{eq4}) increases since $|\alpha|=|\beta| \sum_{i=0}^{Q_a} a_i \sum_{i=0}^{Q_b} c_i$. For the simulations we assume $\mu_a=\mu_b=\mu_c=0.2$. Consequently, as the dyadic-backscatter-link channel gain, i.e., $\alpha$, increases, a better BER performance can be achieved. Note that in general RIS includes larger numbers of reflectors. However in order to show the BER performances in the same frame with the base-line method, we only considered small number of reflectors.

\subsection{Coverage Performance Evaluation} 
\begin{figure}[t]
\centering
\captionsetup{width=1\linewidth}
  \includegraphics[width=9cm, height=6cm]{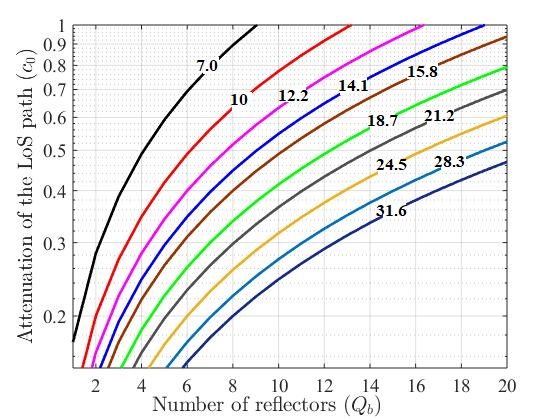}%
  \caption{Coverage enhancement evaluation of the proposed method. As the number of the reflectors on the RIS increases, better coverage enhancement gain is achieved.}%
  \label{fig:5}
\end{figure}
For this set of simulations, the function $Coverage\_ratio()$ is defined, which measures the ratio of the coverage distance in the proposed model to the non-RIS model in the existing AmBC methods. From (\ref{eqdc}), we have $Coverage\_ratio(Q_b,c_0)=\frac{\mathsf{d}_{\text{new}}}{\mathsf{d}_{\text{old}}}=\sqrt{\frac{0.5Q_b^2+Q_b}{c_0^2}}$ for backward-link channel where $\mu_c^2$ and $\sigma_c^2$ both are set to $0.5$ and $1$, respectively. Fig. \ref{fig:5} shows a set of simulation results for different distance ratios. It is shown that when $c_0$ decreases, the proposed method profits from multipath gains reflected from the RIS reflector to enhance the signal power at the reader and consequently enhance the coverage of the communication.

\subsection{Data Transmission Rate and Interference Evaluation} 
\begin{figure}[t]
\centering
\captionsetup{width=1\linewidth}
  \includegraphics[width=8.5cm, height=5.5cm]{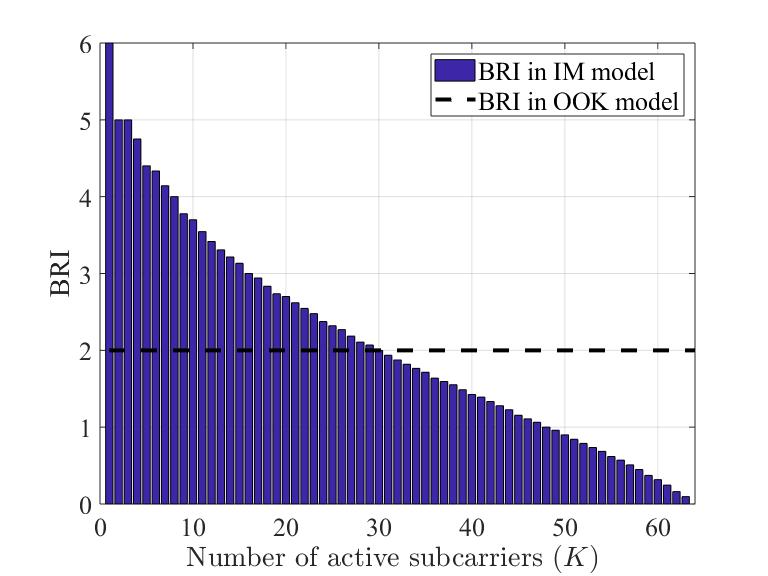}%
  \caption{BRI evaluation of the proposed method in both IM and OOK modulation models.}%
  \label{fig:3}
\end{figure}

In this part, we define a bit-rate-to-interference (BRI) gain function as the bit rate, i.e., $\eta$, over the interference produced by active subcarriers for the original ambient OFDM signal. Let $\lambda$ denote the BRI. With OOK modulation, considering $\eta=N_s$ transmitted bits over an average of $\frac{N_s}{2}$ active subcarriers, the BRI is fixed and given by
\begin{equation}
\lambda_{\text{OOK}}=\frac{\eta}{\frac{N_s}{2} \times {E_b}}= \left.\frac{N_s}{\frac{N_s}{2}\times {E_b}}\right|_{E_b=1}=\, 2.
\end{equation}
The dashed line in Fig. \ref{fig:3} shows the fixed BRI for OOK. However, with IM, from (\ref{eqeta}), we have
\begin{equation}
   \lambda_{\text{IM}}=\left.\frac{\eta}{K \times E_b}\right|_{E_b=1} =\frac{\lfloor \log_2 \binom{N_s}{K} 
\rfloor}{K}. 
\end{equation}

Fig. \ref{fig:3} shows that BRI gain decreases as the number of active subcarriers, i.e., $K$, increases in IM model.

\begin{figure}[t]
\centering
\captionsetup{width=1\linewidth}
  \includegraphics[width=8.5cm, height=5.5cm]{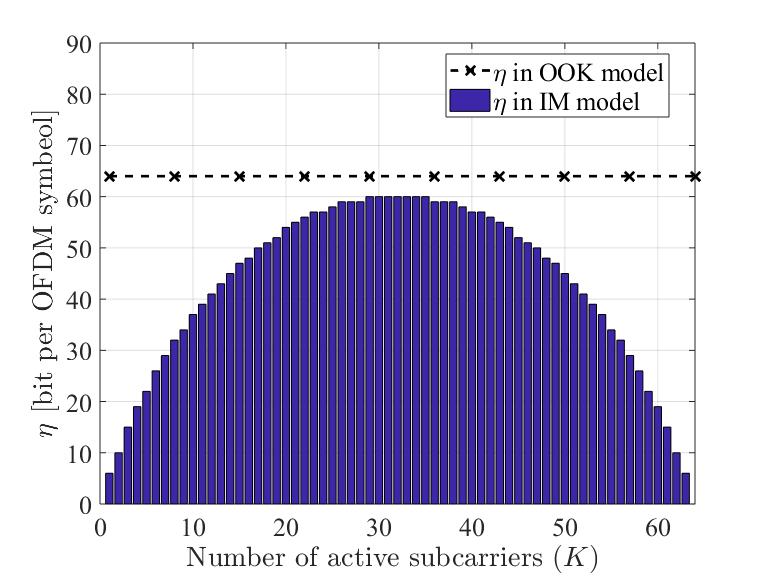}%
  \caption{Bit rate evaluation for both IM and OOK modulation models.}%
  \label{fig:4}
\end{figure}

 Moreover, as shown in Fig. \ref{fig:4}, the OOK model has a fixed bit rate of $\eta=N_s(= 64)$ bits per each OFDM symbol. This fixed bit rate is shown by the dashed line with X markers in Fig. \ref{fig:4}. However, the bit rate of IM model varies as in (\ref{eqeta}). The highest bit rate of IM model, which is $\eta=60$, can be achieved when $K=\frac{N_s}{2}(=32)$.
 

\section{Conclusions}
In this study, we proposed a new AmBC method over ambient OFDM subcarriers in conjunction with RIS to enhance the data rate, coverage and the BER performance for a short-range communication. It transmits one bit per each ambient OFDM subcarrier using OOK modulation and RIS enhances the quality of the signal in rich-scattering environments. Exploiting all squeezed orthogonal OFDM subcarriers supports a higher data rate and RIS improves the BER performance of our proposed model. Moreover, a closed-form optimal decision threshold was derived for the amplitude of the received subcarrier to detect the data at the reader. An analytical expression for the BER was also derived which coincided with the simulation results. From the simulation results, it was shown that the proposed method outperforms the existing method in \cite{of-0} as a baseline method in terms of BER performance and bit rate. Furthermore, IM had also been proposed as a solution to control the power of the signal and reduce the interference for the original OFDM signal.



\bibliographystyle{ieeetr}
\bibliography{ref}
\end{document}